\documentclass[letterpaper,english]{spie}
\usepackage[T1]{fontenc}
\usepackage[latin9]{inputenc}
\usepackage{verbatim}
\usepackage{url}
\usepackage{amsmath}
\usepackage{amssymb}
\usepackage{graphicx}

\makeatletter

\usepackage{url}
\usepackage[pdfpagemode=None,pdfstartview=FitH,pdfview=FitH,colorlinks=true,pdftitle=Perceptual Vector Quantization for Video Coding,pdfauthor=Jean-Marc Valin]{hyperref}

\@ifundefined{showcaptionsetup}{}{%
 \PassOptionsToPackage{caption=false}{subfig}}
\usepackage{subfig}
\makeatother

\usepackage{babel}
\begin{document}

\title{Perceptual Vector Quantization for Video Coding}

\author{Jean-Marc Valin and Timothy B. Terriberry\skiplinehalf Mozilla,
Mountain View, USA\skiplinehalf Xiph.Org Foundation\authorinfo{Copyright 2014-2015 Mozilla Foundation. This work is licensed under \href{https://creativecommons.org/licenses/by/3.0/}{CC-BY 3.0}.\\Send correspondence to Jean-Marc Valin <\href{mailto:jmvalin@jmvalin.ca}{jmvalin@jmvalin.ca}>.}}
\maketitle
\begin{abstract}
This paper applies energy conservation principles to the Daala video
codec using gain-shape vector quantization to encode a vector of AC
coefficients as a length (gain) and direction (shape). The technique
originates from the CELT mode of the Opus audio codec, where it is
used to conserve the spectral envelope of an audio signal. Conserving
energy in video has the potential to preserve textures rather than
low-passing them. Explicitly quantizing a gain allows a simple contrast
masking model with no signaling cost. Vector quantizing the shape
keeps the number of degrees of freedom the same as scalar quantization,
avoiding redundancy in the representation. We demonstrate how to predict
the vector by transforming the space it is encoded in, rather than
subtracting off the predictor, which would make energy conservation
impossible. We also derive an encoding of the vector-quantized codewords
that takes advantage of their non-uniform distribution. We show that
the resulting technique outperforms scalar quantization by an average
of 0.90~dB on still images, equivalent to a 24.8\% reduction in bitrate
at equal quality, while for videos, the improvement averages 0.83~dB,
equivalent to a 13.7\% reduction in bitrate.
\end{abstract}

\section{Introduction}

Video codecs are traditionally based on scalar quantization of discrete
cosine transform (DCT) coefficients with uniform quantization step
sizes. High-quality encoders signal quantization step size changes
at the macroblock level to account for contrast masking~\cite{Osberger1999}.
The adjustment cannot be applied at a level smaller than a full macroblock
and applies to all frequencies uniformly. Audio codecs have long considered
frequency-dependent masking effects and more recently, the Opus codec~\cite{rfc6716}
has integrated masking properties as part of its bitstream in a way
that does not require explicit signaling. 

We apply the same principles to the Daala video codec~\cite{DaalaWebsite},
using gain-shape vector quantization to conserve energy. We represent
each vector of AC coefficients by a length (gain) and a direction
(shape). The \emph{shape} is an $N$-dimensional vector that represents
the coefficients after dividing out the gain. The technique originates
from the CELT mode~\cite{Valin2013} of the Opus audio codec, where
splitting the spectrum into bands and coding an explicit gain for
each band preserves the spectral envelope of the audio. Similarly,
conserving energy in videohas the potential to preserve textures rather
than low-passing them. Explicitly quantizing a gain allows a simple
contrast masking model with no signaling cost. Vector quantizing the
shape keeps the number of degrees of freedom the same as scalar quantization,
avoiding redundancy in the representation.

Unlike audio, video generally has good predictors available for the
coefficients it is going to code, e.g., the motion-compensated reference
frame. The main challenge of gain-shape quantization for video is
figuring out how to apply that prediction without losing the energy-conserving
properties of explicitly coding the gain of the input vector. We cannot
simply subtract the predictor from the input vector: the length of
the resulting vector would be unrelated to the gain of the input.
That is, the probability distribution of the points on the hypersphere
is highly skewed by our predictor, rather than being fixed, like in
CELT. Section~\ref{sec:Prediction} shows how to transform the hypersphere
in a way that allows us to easily model this skew. Finally, we derive
a new way of encoding the shape vector that accounts for the fact
that low-frequency AC coefficients have a larger variance than high-frequency
AC coefficients. We then compare the quality obtained using the proposed
technique to scalar quantization on both still images and videos.

\section{Gain-Shape Quantization}

Most video and still image codecs use scalar quantization. That is,
once the image is transformed into frequency-domain coefficients,
they quantize each (scalar) coefficient separately to an integer index.
Each index represents a certain scalar value that the decoder uses
to reconstruct the image. For example, using a quantization step size
of 10, a decoder receiving indices -2, +4, produces reconstructed
coefficients of -20, +40, respectively. Scalar quantization has the
advantage that each dimension can be quantized independently from
the others. 

With vector quantization, an index no longer represents a scalar value,
but an entire vector. The possible reconstructed vectors are no longer
constrained to the regular rectangular grid above, but can be arranged
in any way we like, either regularly, or irregularly. Each possible
vector is called a codeword and the set of all possible codewords
is called a codebook. The codebook may be finite if we have an exhaustive
list of all possibilities, or it can also be infinite in cases where
it is defined algebraically (e.g., scalar quantization is equivalent
to vector quantization with a particular infinite codebook).

Gain-shape vector quantization represents a vector by separating it
into a length and a direction. The \emph{gain} (the length) is a scalar
that represents how much energy is in the vector. The \emph{shape}
(the direction) is a unit-norm vector that represents how that energy
is distributed in the vector. In two dimensions, one can construct
a gain-shape quantizer using polar coordinates, with gain being the
radius and the shape being the angle. In three dimensions, one option
(of many) is to use latitude and longitude for the shape. In Daala,
we don't restrict ourselves to just two or three dimensions. For a
4x4 transform, we have 16~DCT coefficients. We always code the DC
separately, so we are left with a vector of 15 AC coefficients to
quantize. If we quantized the shape vector with scalar quantization,
we would have to code 16 values to represent these 15 coefficients:
the 15 normalized coefficients plus the gain. That would be redundant
and inefficient. But by using vector quantization, we can code the
direction of an $N$-dimensional vector with $(N-1)$ degrees of freedom,
just like polar coordinates in 2-D and latitude and longitude in 3-D.
Classically, vector quantization provides other advantages, such as
the \emph{space filling advantage}, the \emph{shape advantage}, and
the \emph{memory advantage}~\cite{LG89}, but we benefit very little,
if at all, from these. Our primary interest is in elminating redundancy
in the encoding.

In Daala, as in CELT, we use a normalized version of the pyramid vector
quantizer described by Fischer~\cite{Fisher1986} for the shape codebook.
Fischer's codebook is defined as the set of all $N$-dimensional vectors
of integers where the absolute value of the integers sum to $K$:
\begin{equation}
\mathbf{y}\in\mathbb{Z}^{N}:\sum_{i=0}^{N-1}\left|y_{i}\right|=K\ .\label{eq:pvq-codebook-def}
\end{equation}
$K$ is called the number of pulses because one can think of building
such a vector by distributing $K$ different pulses of unit amplitude,
one at a time, among the $N$ dimensions. The pulses can point in
either the positive or negative direction, but all of the ones in
the same dimension must point in the same direction. For $N=2$ and
$K=2$, our codebook has 8 entries:
\[
(2,0),\ (-2,0),\ (1,1),\ (1,-1),\ (-1,1),\ (-1,-1),\ (0,2),\ (0,-2)
\]

The normalized version of the codebook simply normalizes each vector
to have a length of 1. If $\mathbf{y}$ is a codeword in Fischer's
non-normalized codebook, then the corresponding normalized vector
is simply 
\begin{equation}
\mathbf{u}=\mathbf{y}/\left\Vert \mathbf{y}\right\Vert _{L2}\ .
\end{equation}

The advantages of this codebook are that it is flexible (any value
of $K$ will work), it does not require large lookup tables for each
$K$, and it is easy to search. Its regularity means that we can construct
vectors on the fly and don't have to examine every vector in the codebook
to find the best representation for our input.

\begin{figure}
\centering{\includegraphics[width=0.7\columnwidth]{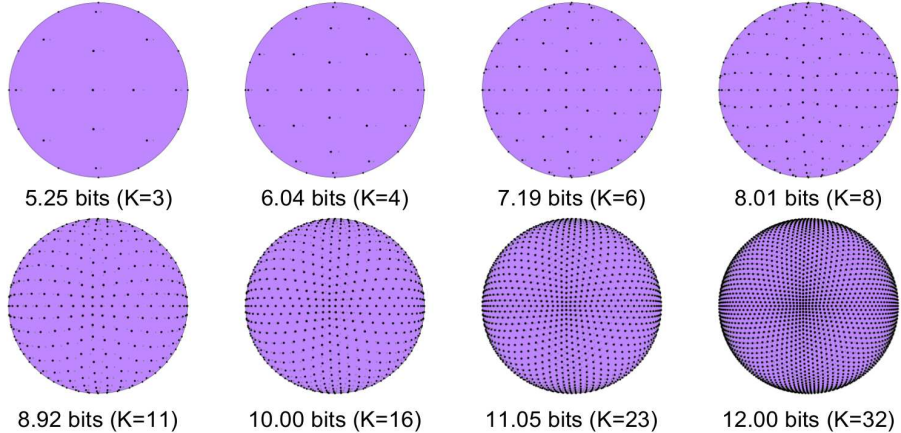}}

\caption{Normalized-pyramid vector quantizer in three dimensions. The codewords
are spread relatively uniformly on the sphere (perfect uniformity
is provably impossible in 3 dimensions, and not actually desirable
for Laplacian sources).}
\end{figure}

The number of dimensions, $N$, is fixed, so the parameter $K$ determines
the quantization resolution of the codebook (i.e., the spacing between
codewords). As the gain of our vector increases, we need more points
on our $N$-dimensional sphere to keep the same quantization resolution,
so $K$ must increase as a function of the gain. 

For the gain, we can use simple scalar quantization:
\begin{equation}
\hat{g}=\gamma Q\ ,
\end{equation}
where $Q$ is the quantization step size, $\gamma$ is the quantization
index, and $\hat{g}$ is the reconstructed (quantized) gain.

\section{Prediction}

\label{sec:Prediction}Video codecs rarely code blocks \emph{from
scratch}. Instead, they use inter- or intra-prediction. In most cases,
that prediction turns out to be very good, so we want to make full
use of it. The naive approach is to just subtract the prediction from
the coefficients being coded, and then apply gain-shape quantization
to the residual. While this can work, we lose some of the benefits
of gain-shape quantization because the gain no longer represents the
contrast of the block being coded, only how much the image changed
compared to the prediction. 

\begin{figure}
\centering{\includegraphics[width=0.4\columnwidth]{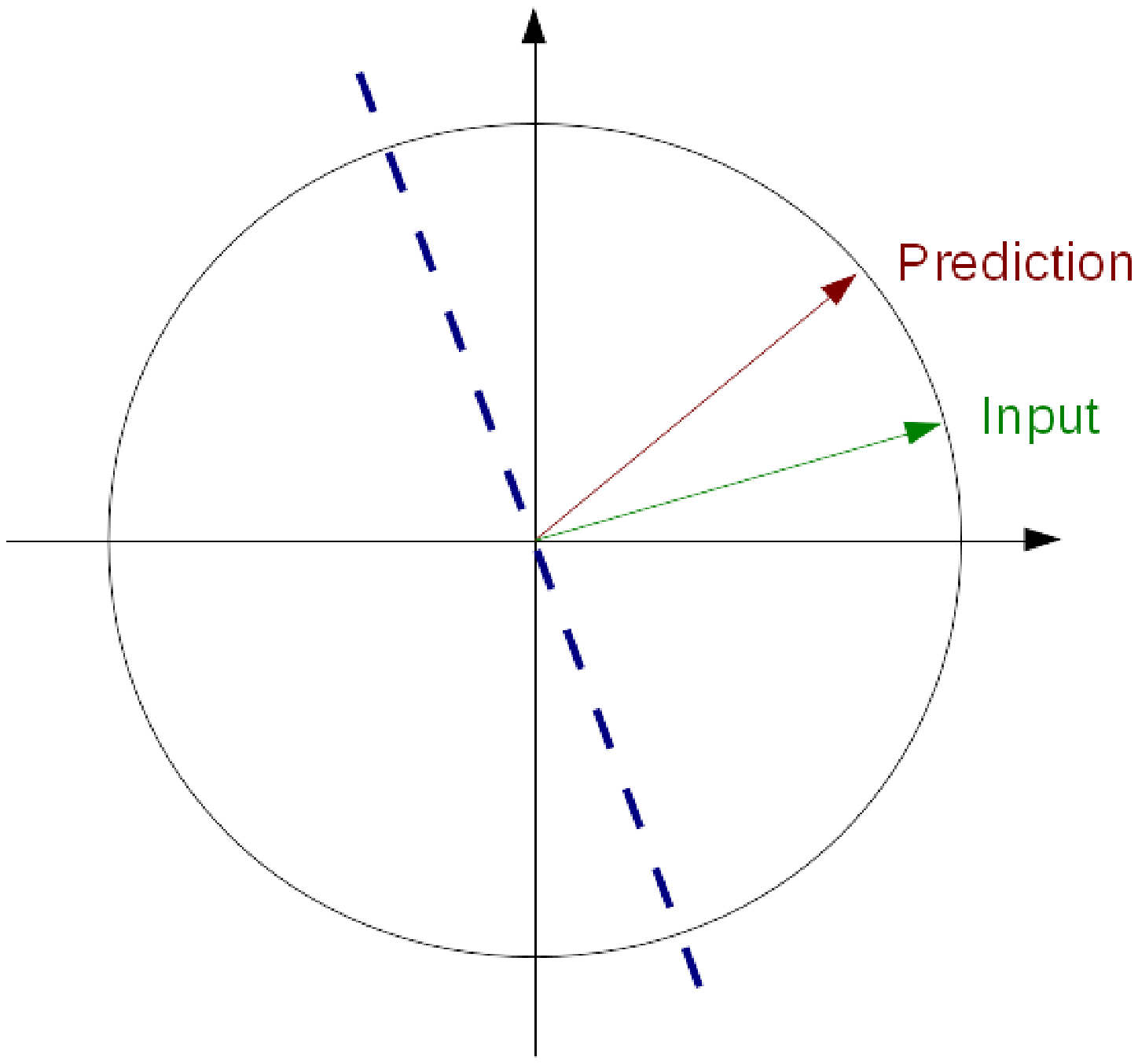}\hspace{1em}\includegraphics[width=0.4\columnwidth]{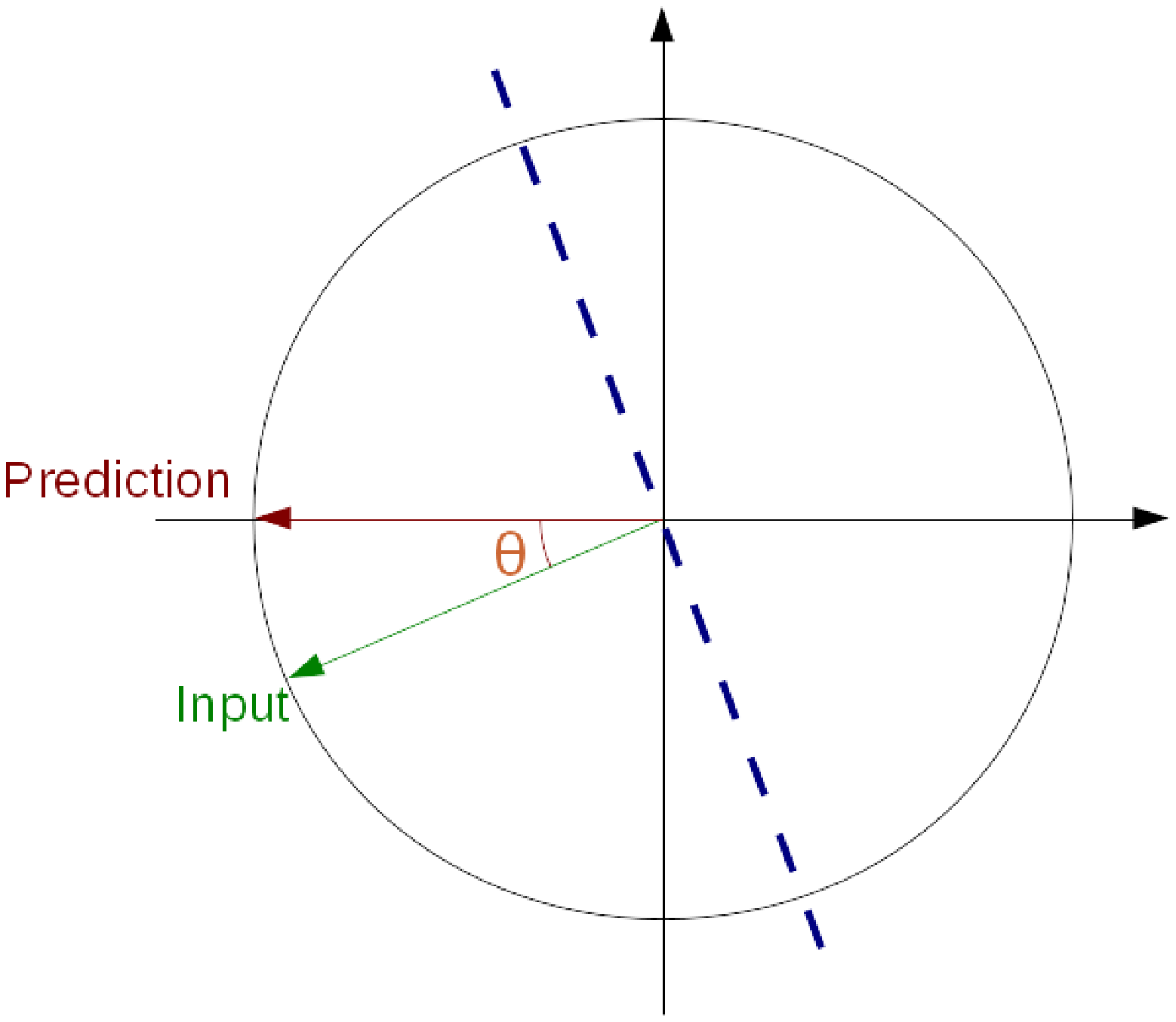}}\caption{Householder reflection in 2~dimensions. Left: original prediction
and input $\mathbf{x}$. Right: reflected input $\mathbf{z}$.\label{fig:Householder-reflection-steps}}
\end{figure}

An early attempt at warping the shape codebook~\cite{Valin2010}
to increase resolution close to the prediction gave little benefit
despite the increase in complexity. In this work, we instead transform
the space so that the predictor lies along one of the axes, allowing
us to treat that axis as ``special''. This gives us much more direct
control over the quantization and probability modeling of codewords
close to the predictor than the codebook warping approach. We compute
the transform using a Householder reflection, which is also computationally
cheaper than codebook warping. The Householder reflection constructs
a reflection plane that turns the prediction into a vector with only
one non-zero component. Let $\mathbf{r}$ be the vector containing
the prediction coefficients. The Householder reflection is defined
by a vector that is normal to the reflection plane:
\begin{equation}
\mathbf{v}=\frac{\mathbf{r}}{\left\Vert \mathbf{r}\right\Vert }+s\mathbf{e}_{m}\,,\label{eq:reflection-vector}
\end{equation}
where $\mathbf{e}_{m}$ is a unit vector along axis $m$, $s=\mathrm{sign\left(r_{m}\right)}$
and axis $m$ is selected based on the largest component of the prediction,
to minimize numerical error. The decoder can compute both values without
sending any additional side information. The input vector $\mathbf{x}$
is reflected using $\mathbf{v}$ by computing
\begin{equation}
\mathbf{z}=\mathbf{x}-2\frac{\mathbf{v}^{T}\mathbf{x}}{\mathbf{v}^{T}\mathbf{v}}\mathbf{v}\ .\label{eq:Householder-computation}
\end{equation}
Fig.~\ref{fig:Householder-reflection-steps} illustrates the process.
From this point, we focus on quantizing $\mathbf{z}$, knowing that
its direction is likely to be close to $\mathbf{e}_{m}$.

Once the prediction is aligned along an axis, we can code how well
it matches the input vector using an angle $\theta$, with
\begin{equation}
\cos\theta=\frac{\mathbf{x}^{T}\mathbf{r}}{\left\Vert \mathbf{x}\right\Vert \left\Vert \mathbf{r}\right\Vert }=\frac{\mathbf{z}^{T}\mathbf{r}}{\left\Vert \mathbf{z}\right\Vert \left\Vert \mathbf{r}\right\Vert }=-\frac{sz_{m}}{\left\Vert \mathbf{z}\right\Vert }\label{eq:cosine-distance-theta}
\end{equation}

The position of the input vector along the prediction axis on the
unit hypersphere is simply $\cos\theta$, and the distance between
the input and the prediction axis is $\sin\theta$. We could choose
to code $\cos\theta$, or $1-\cos\theta$, or $\sin\theta$, but directly
coding $\theta$ using scalar quantization turns out to be the optimal
choice for minimizing mean-squared error (MSE).We then apply gain-shape
quantization to the remaining dimensions, knowing that the gain of
the remaining coefficients is just the original gain multiplied by
$\sin\theta$. The quantized coefficients are reconstructed from the
gain $\hat{g}$, the angle $\hat{\theta}$, and a unit-norm codeword
$\mathbf{u}$ as
\begin{equation}
\mathbf{\hat{z}}=\hat{g}\left(\cos\hat{\theta}\mathbf{e}_{m}+\sin\hat{\theta}\mathbf{u}\right)\ ,\label{eq:reconstructed-normalized-vector}
\end{equation}
where $\hat{g}$, $\hat{\theta}$, and $\mathbf{u}$ are jointly rate-distortion-optimized
in the encoder. Because of the angle $\hat{\theta}$, dimension $m$
can be omitted from $\mathbf{u}$, which now has $N-1$ dimensions
and $N-2$ degrees of freedom (since its norm is unity). Thus the
total number of degrees of freedom we have to encode for an $N$-dimensional
vector remains $N$. Although one might expect the Householder reflection
to distort the statistics of the coefficients in the $N-1$ dimensions
remaining in $\mathbf{u}$ in unpredictable ways, since high-variance
LF coefficients could be mixed with low-variance HF coefficients,
in practice we find they remain relatively unchanged.

\section{Activity Masking}

Video and still image codecs have long taken advantage of the fact
that contrast sensitivity of the human visual system depends on the
spatial frequency of a pattern. The contrast sensitivity function
is to vision what the absolute threshold of hearing is to audition.

Another factor that influences our perception of contrast is the presence
of more contrast at the same location. We call this activity masking.
It is the visual equivalent of auditory masking, where the presence
of a sound can mask another sound. While auditory masking has been
at the center of all audio coding technology since the 80s~\cite{johnston1988transform},
the use of activity masking in video coding is comparatively new and
less advanced. This is likely due to the fact that activity masking
effects are much less understood than auditory masking, which has
been the subject of experiments since at least the 30s~\cite{fletcher1937relation}.

\begin{figure}
\centering{\includegraphics[width=0.8\columnwidth]{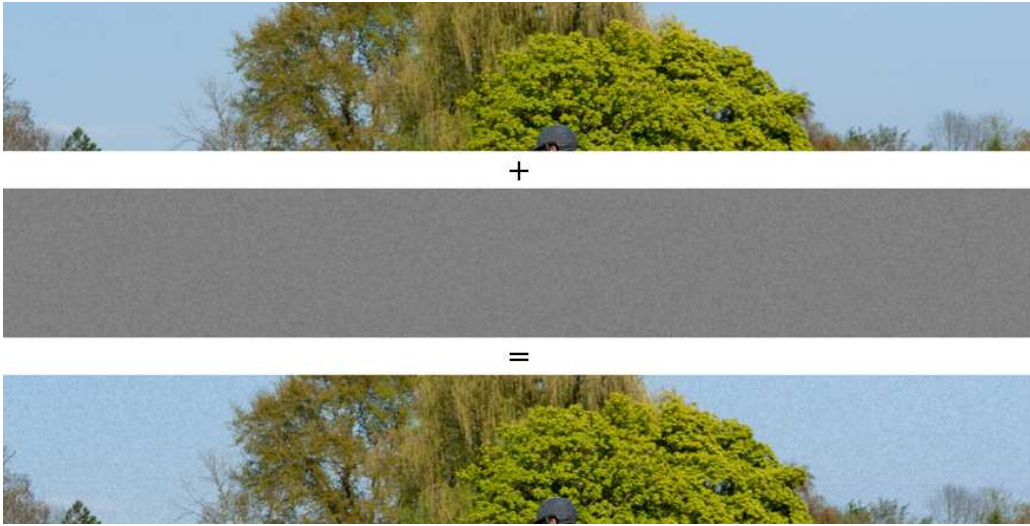}}\caption{Contrast-dependent masking of noise. Top: original image. Middle:
spatially uniform noise. Bottom: noisy image. The noise added to the
bottom image is clearly visible in the sky area because of the lack
of texture in that part of the original image. On the other hand,
it is barely visible in the trees because it is masked by the texture
present in the original image.\label{fig:Contrast-dependent-masking-demo}}
\end{figure}

Fig.~\ref{fig:Contrast-dependent-masking-demo} demonstrates activity
masking and how contrast in the original image affects our perception
of noise. Taking into account activity masking has two related, but
separate goals:
\begin{itemize}
\item Increasing the amount of quantization noise in high-contrast regions
of the image, while decreasing it in low-contrast regions; and
\item Ensuring that low-contrast regions of the image do not get completely
washed out (i.e. keeping some AC).
\end{itemize}
Together, these two tend to make bit allocation more uniform throughout
the frame, though not completely uniform \textemdash{} unlike CELT,
which has uniform allocation due to the logarithmic sensitivity of
the ear.

\begin{figure}
\centering{\includegraphics[width=0.5\columnwidth]{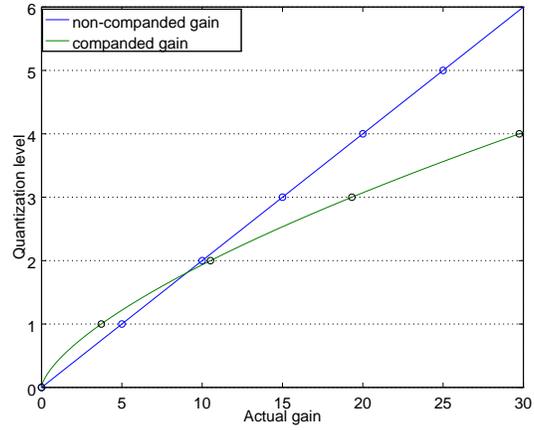}}\caption{Gain companding example for $Q=5$. For small gains, the quantization
levels result in a small change to the gain, but for large gains,
the the interval is larger.\label{fig:Gain-companding}}
\end{figure}

\begin{figure}
\centering{\includegraphics[width=0.35\columnwidth]{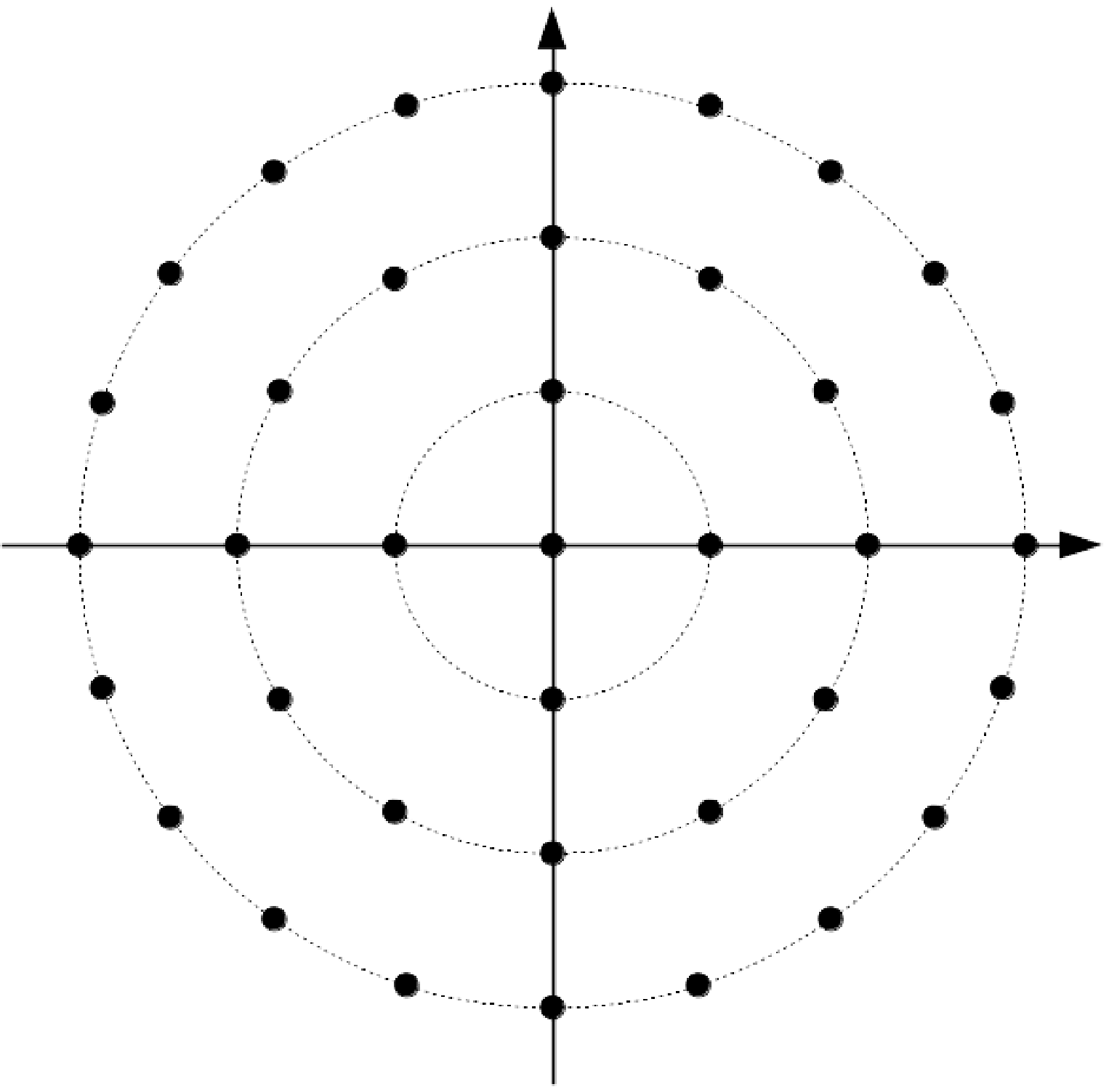}\hspace{1em}\includegraphics[width=0.35\columnwidth]{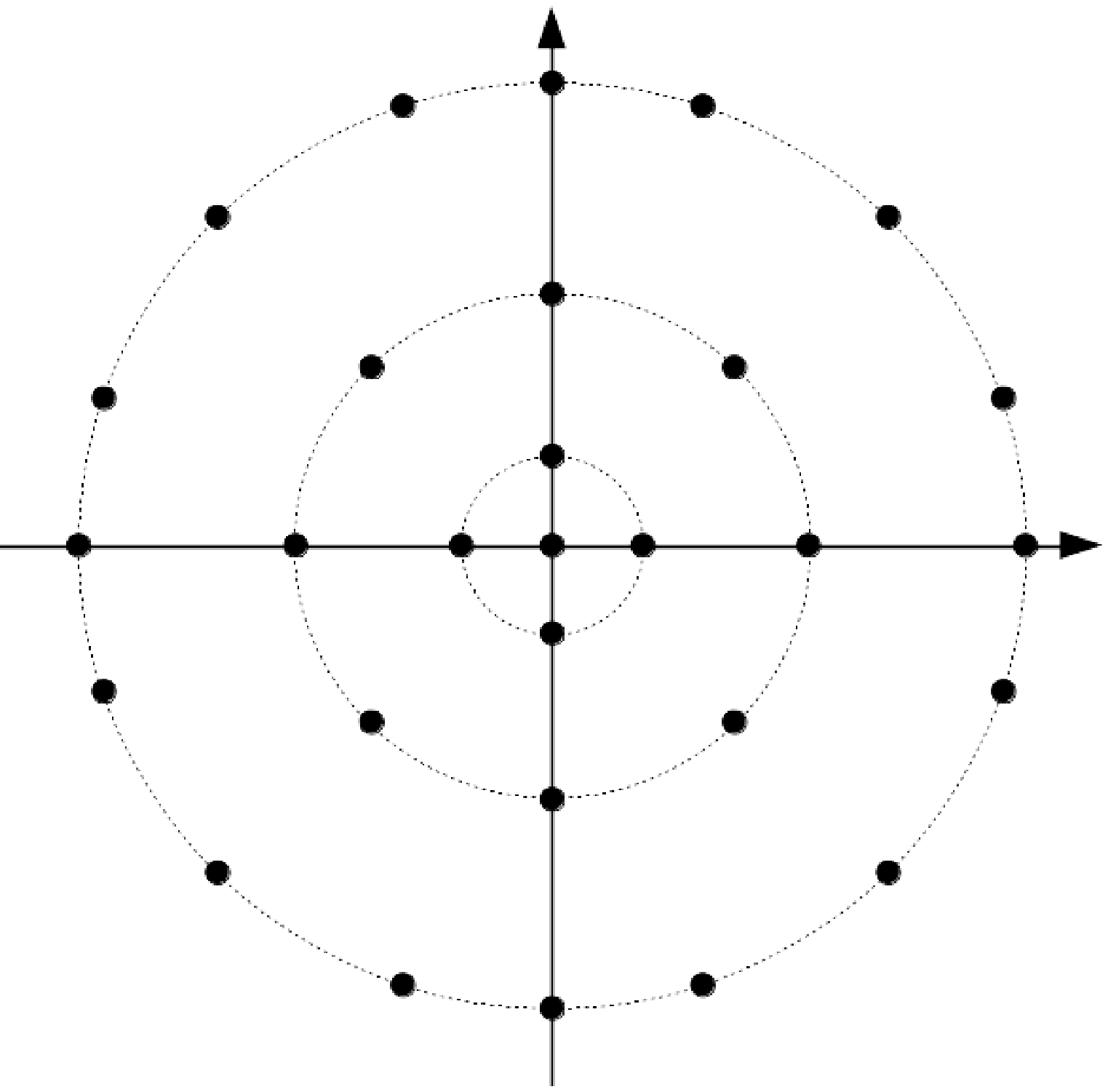}}\caption{Effect of activity masking on the codebook. Left: no activity masking
($\alpha=0$). Right: with activity masking ($\alpha=1/3$).\label{fig:Effect-activity-masking-codebook}}
\end{figure}

We want to make the quantization noise vary as a function of the original
contrast, typically using a power law. In Daala, we make the quantization
noise follow the amount of contrast raised to the power $\alpha=1/3$.
We can control the resolution of the shape quantization simply by
controlling the value of $K$ as a function of the gain. For gain
quantization we can derive a scalar quantizer based on quantization
index $\gamma$ that follows the desired resolution:
\begin{align}
\frac{\partial\hat{g}}{\partial\gamma} & =Q\hat{g}^{\alpha}\nonumber \\
\hat{g}^{-\alpha}\partial\hat{g} & =Q\partial\gamma\nonumber \\
\frac{\hat{g}^{1-\alpha}}{1-\alpha} & =Q\gamma\nonumber \\
\hat{g} & =\left(\left(1-\alpha\right)Q\gamma\right)^{1/\left(1-\alpha\right)}\nonumber \\
\hat{g} & =Q_{g}\gamma^{1/\left(1-\alpha\right)}\\
\hat{g} & =Q_{g}\gamma^{\beta}\label{eq:gain-scalar-quantization}
\end{align}
where $Q_{g}=\left(\left(1-\alpha\right)Q\right)^{\beta}$ is the
new gain resolution and ``master'' quality parameter and $\beta=1/\left(1-\alpha\right)$
controls the activity masking. The effect of activity masking on our
gain quantizer is equivalent to a uniform scalar quantizer applied
to the companded gain. The encoder compands the gain by raising it
to the power $\left(1-\alpha\right)$, and the decoder expands it
back by raising it to the power $1/\left(1-\alpha\right)$. This results
in a better resolution close to $g=0$ and worse resolution as the
gain increases, as shown in Fig.~\ref{fig:Gain-companding}. Fig.~\ref{fig:Effect-activity-masking-codebook}
shows the effect of activity masking on the spherical codebook.

When using prediction, it is important to apply the gain companding
to the original $N$-dimensional coefficient vector rather than to
the $N-1$ remaining coefficients after prediction. In practice, we
only want to consider activity masking in blocks that have texture
rather than edges. The easiest way to do this is to assume all 4x4
blocks have edges and all 8x8 to 32x32 blocks as having texture. Also,
we currently only use activity masking for luma because chroma masking
is much less well-understood. For blocks where we disable activity
masking, we still use the same quantization described above, but set
$\alpha=0$.

\section{Coding Resolution}

It is desirable for a single quality parameter to control $K$ and
the resolution of the gain and $\theta$. That quality parameter should
also take into account activity masking. The MSE-optimal angle quantization
resolution is the one that matches the gain resolution. Assuming the
vector distance is approximately equal to the arc distance, we have
\begin{equation}
Q_{\theta}=\frac{\partial\hat{g}/\partial\gamma}{\hat{g}}=\frac{Q_{g}\beta\gamma^{\beta-1}}{Q_{g}\gamma^{\beta}}=\frac{\beta}{\gamma}\ .\label{eq:theta-quantization-step}
\end{equation}
The quantized angle is thus given by
\begin{equation}
\hat{\theta}=Q_{\theta}\tau\ ,\label{eq:theta-dequant}
\end{equation}
where $\tau$ is the angle quantization index. In practice, we round
$Q_{\theta}$ to an integer fraction of $\pi/2$.

\subsection{Setting $K$}

\begin{figure}
\centering{\includegraphics[width=0.6\columnwidth]{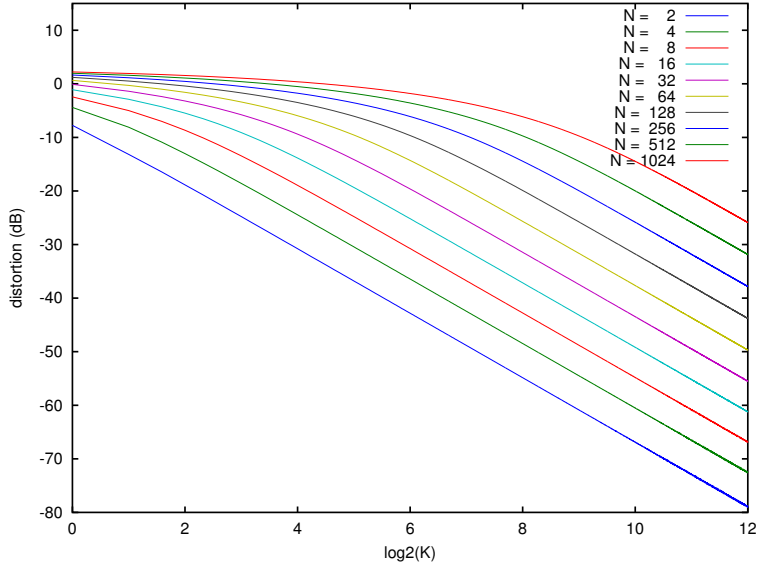}}

\caption{PVQ distortion as a function of $N$ and $K$ for a Laplace-distributed
source\label{fig:PVQ-distortion-N-K}}
\end{figure}

Using an \emph{i.i.d.} Laplace source normalized to unit norm, we
simulated quantization with different values of $N$ and $K$. The
resulting distortion is shown in Fig.~\ref{fig:PVQ-distortion-N-K}.
The asymptotic distortion for large values of $K$ is approximately
\begin{equation}
D_{pvq}=\frac{\left(N-1\right)^{2}+C_{K}\left(N-1\right)}{24K^{2}}\ ,
\end{equation}
with $C_{K}=4.$

The distortion due to scalar quantization of the gain is (asymptotically)
\begin{align}
D_{g} & =\frac{1}{12}\left(\partial\hat{g}/\partial\gamma\right)^{2}\nonumber \\
 & =\frac{\beta^{2}Q_{g}^{2}\gamma^{2\beta-2}}{12}\ .
\end{align}
To achieve uniform distortion along all dimensions, the distortion
due to the $N-2$ PVQ degrees of freedom must be $N-2$ times greater
than that due to quantizing the gain, so

\begin{align}
\left(N-2\right)D_{g} & =\left(\hat{g}\sin\hat{\theta}\right)^{2}D_{pvq}\nonumber \\
\frac{\left(N-2\right)\beta^{2}Q_{g}^{2}\gamma^{2\beta-2}}{12} & =\left(Q_{g}\gamma^{\beta}\sin\hat{\theta}\right)^{2}\frac{\left(N-2\right)^{2}+C_{K}\left(N-2\right)}{24K^{2}}\nonumber \\
\left(N-2\right)\beta^{2} & =\frac{\gamma^{2}\sin^{2}\hat{\theta}\left[\left(N-2\right)^{2}+C_{K}\left(N-2\right)\right]}{2K^{2}}\nonumber \\
K & =\frac{\gamma\sin\hat{\theta}}{\beta}\sqrt{\frac{N+C_{K}-2}{2}}\label{eq:K-res}
\end{align}
This avoids having to signal $K$. It is determined only by $\gamma$
and $\hat{\theta}$, both known by the decoder.

\subsection{Gain Prediction and Loss Robustness}

When using inter-frame prediction, the gain of the predictor is likely
to be highly correlated with the gain we are trying to code. We want
to use it to save on the cost of coding $\hat{g}$. However, since
(\ref{eq:K-res}) depends on the quantized gain $\gamma$, any encoder-decoder
mismatch in the predicted gain (e.g., caused by packet loss) would
cause the decoder to decode the wrong number of symbols, causing an
encoder-decoder mismatch in the entropy coding state and making the
rest of the frame unrecoverable. For video applications involving
unreliable transport (such as RTP), this must be avoided. 

Since the prediction is usually good, we can approximate $\sin\hat{\theta}\approx\hat{\theta}$.
Substituting (\ref{eq:theta-dequant}) into (\ref{eq:K-res}), we
have
\begin{align}
K & =\frac{\gamma\sin\frac{\beta\tau}{\gamma}}{\beta}\sqrt{\frac{N+C_{K}-2}{2}}\nonumber \\
 & \approx\tau\sqrt{\frac{N+C_{K}-2}{2}}\ .\label{eq:robust-K}
\end{align}
Using (\ref{eq:robust-K}), the decoder's value of $K$ will depend
on the quantization index $\tau$, but not the gain, nor the actual
angle $\hat{\theta}$. Any error in the prediction caused by packet
loss may result in different values for $\hat{g}$, $\hat{\theta}$,
or the Householder reflection parameters, but it will not cause entropy
decoding errors, so the effect will be localized, and the frame will
still be decodable.

\section{Application to Video Coding}

\begin{figure}[t]
\centering{\includegraphics[width=0.32\columnwidth]{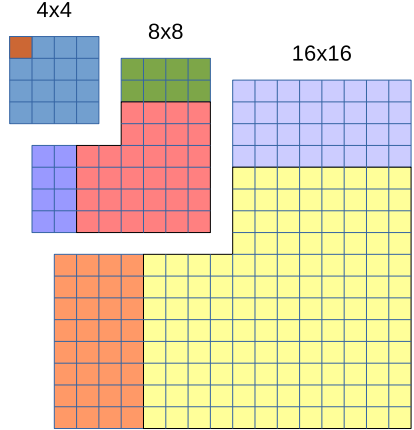}}

\caption{Band definition for 4x4, 8x8 and 16x16 blocks. The low frequencies
are recursively divided following the pattern of smaller blocks. 4x4
blocks have all their AC coefficients in a single band.\label{fig:Band-layout}}
\end{figure}

To apply gain-shape vector quantization to DCT coefficients, it is
important to first divide the coefficients into \emph{frequency bands}
just like for audio, to avoid moving energy across octaves or orientations
during the quantization process. Fig.~\ref{fig:Band-layout} illustrates
the current bands we use for different block sizes. Blocks of 4x4,
8x8 and 16x16 are split into 1, 4, and 7~bands, respectively. DC
is always excluded and scalar quantized separately. 

\subsection{Signaling}

The last thing we need to discuss is how the parameters are transmitted
from the encoder to the decoder. An important aspect of PVQ is to
build activity masking into the codec, so that no extra information
needs to be transmitted. We avoid explicitly signaling quantization
resolution changes depending on the amount of contrast like other
codecs currently do. Better, instead of only being able to signal
activity masking rate changes at the macroblock level, we can do it
at the block level and even per frequency band. We can treat horizontal
coefficients independently of the vertical and diagonal ones, and
do that separately for each octave.

Let's start with the simplest case, where we do not have a predictor.
The first parameter we code is the gain, which is raised to the power
$(1-\alpha)$ when activity masking is enabled. From the gain alone,
the bit-stream defines how to obtain $K$, so the decoder knows what
$K$ is without the need to signal it. Once we know K, we can code
the N-dimensional pyramid VQ codeword. Although it may look like we
code more information than scalar quantization because we code the
gain, it is not the case because knowing K makes it cheaper to code
the remaining vector since we know the sum of the absolute values.

The case with prediction is only slightly more complicated. We still
have to encode a gain, but this time we have a prediction, so we can
use it to make coding the gain cheaper. After the gain, we code the
angle $\theta$ with a resolution that depends on the gain. Last,
we code a VQ codeword in $N-1$ dimensions. We only allow angles smaller
than $\pi/2$, indicating a positive correlation with the predictor.
To handle the cases where the input is poorly or negatively correlated
with the predictor, we code a ``no reference'' flag. When set, we
ignore the reference and only code the gain and an N-dimensional pyramid
VQ codeword (no angle).

\subsection{Coefficient Encoding}

Encoding coefficients quantized with PVQ differs from encoding scalar-quantized
coefficients since the sum of the coefficients' magnitude is known
(equal to $K$). While knowing $K$ places hard limits on the coefficient
magnitudes that reduce the possible symbols that can be coded, much
larger gains can be had by using it to improve probability modeling.
It is possible to take advantage of the known $K$ value either through
modeling the distribution of coefficient magnitudes or by modeling
the length of zero runs. In the case of magnitude modeling, the expectation
of the magnitude of coefficient $n$ is modeled as
\begin{equation}
E\left(\left|y_{n}\right|\right)=\mu\frac{K_{n}}{N-n}\ ,\label{eq:magnitude-encoding}
\end{equation}
where $K_{n}$ is the number of pulses left after encoding coefficients
from $0$ to $n-1$ and $\mu$ depends on the distribution of the
coefficients. For run-length modeling, the expectation of the position
of the next non-zero coefficient is given by
\begin{equation}
E\left(run\right)=\nu\frac{N-n}{K_{n}}\ ,\label{eq:run-length-encoding}
\end{equation}
where $\nu$ also models the coefficient distribution. Given the expectation,
we code a value by assuming a Laplace distribution with that expectation.
The parameters $\mu$ and $\nu$ are learned adaptively during the
coding process, and we can switch between the two strategies based
on $K_{n}$. Currently we start with magnitude modeling and switch
to run-length modeling once $K_{n}$ drops to 1.

\section{Results}

\begin{figure}[t]
\subfloat[Still image, PSNR of scalar vs vector quantization\label{fig:Still-image-flat}]{\includegraphics[width=0.5\columnwidth]{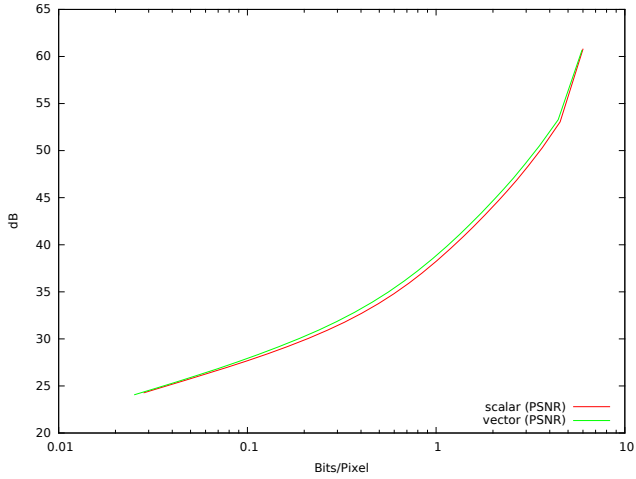}

}\subfloat[Still image, FAST-SSIM with and without masking\label{fig:Still-image-fastssim}]{\includegraphics[width=0.5\columnwidth]{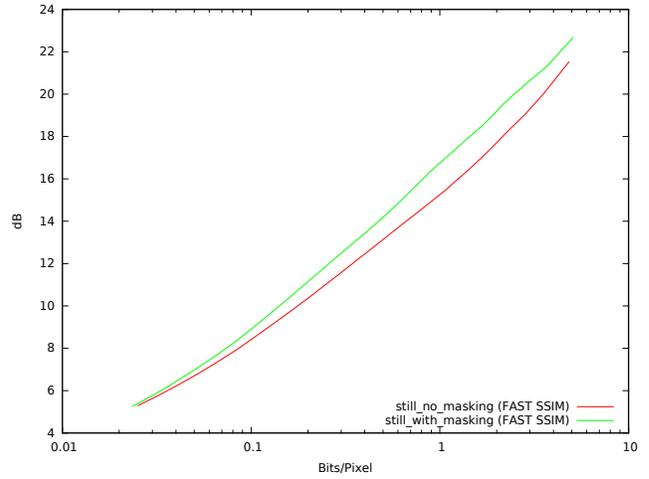}

}

\subfloat[Video, PSNR of scalar vs vector quantization\label{fig:Video-flat}]{\includegraphics[width=0.5\columnwidth]{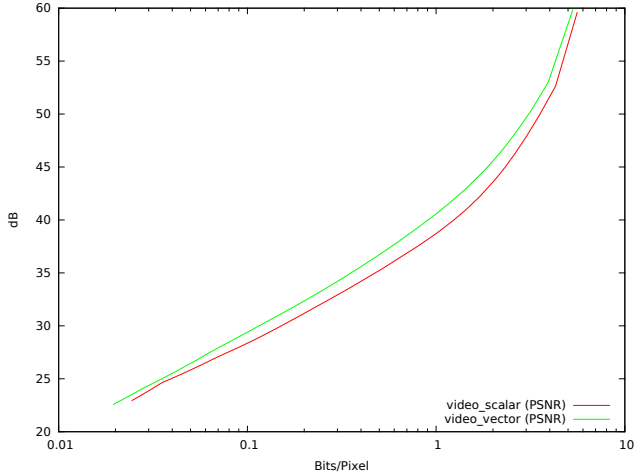}

}\subfloat[Video, FAST-SSIM with and without masking\label{fig:Video-fastssim}]{\includegraphics[width=0.5\columnwidth]{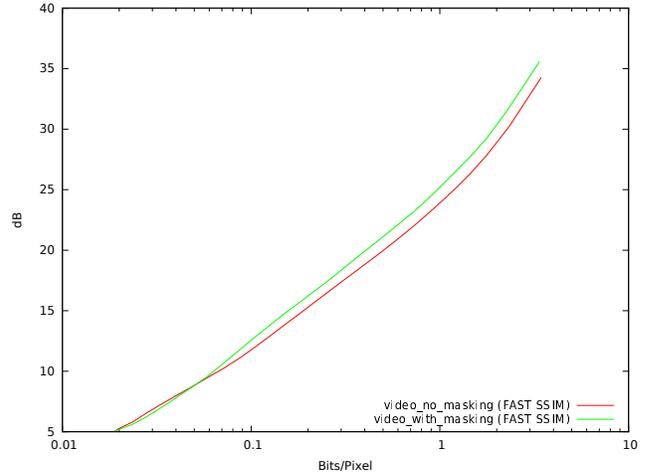}

}

\caption{Comparing scalar quantization to vector quantization with and without
contrast masking.\label{fig:fastssim}}
\end{figure}

The contrast masking algorithm is evaluated on both still a set of
50~still images taken from Wikipedia and downsampled to 1~megapixel\footnote{\url{https://people.xiph.org/~tterribe/daala/subset1-y4m.tar.gz}},
and on a set of short video clips ranging from CIF to 720p in resolution.
First, the PSNR performance of scalar quantization vs. vector quantization
is compared in Fig.~\ref{fig:Still-image-flat} and \ref{fig:Video-flat}.
To make comparison easier, we use a flat quantization matrix for both
scalar and vector quantization. 

Since we expect the use of contrast masking to make measurements such
as PSNR worse, we evaluate its effect using a fast implementation
of multi-scale structural similarity (FAST-SSIM)~\cite{Chen2010}.
Fig.~\ref{fig:Still-image-fastssim} and~\ref{fig:Video-fastssim}
show the FAST-SSIM results with and without contrast masking. For
still images, the average improvement is 0.90~dB, equivalent to a
24.8\% reduction in bitrate at equal quality, while for videos, the
average improvement is 0.83~dB, equivalent to a 13.7\% reduction
in bitrate.

\section{Conclusion}

We have presented a perceptual vector quantization technique for still
images and video. We have shown that it can be used to implement adaptive
quantization based on contrast masking without any signaling in a
way that improves quality. For now, contrast masking is restricted
to the luma planes. It remains to be seen if a similar technique can
be applied to the chroma planes.

This work is part of the Daala project~\cite{DaalaWebsite}. The
full source code, including all of the PVQ work described in this
paper is available in the project git repository~\cite{DaalaGit}.

\bibliographystyle{spiebib}
\bibliography{daala}

\begin{thebibliography}{10}

\bibitem{Osberger1999}
Osberger, W.,  [{\em Perceptual Vision Models for Picture Quality Assessmnet
  and Compression Applications}{\nolinebreak\hspace{0.1em}]}, Queensland
  University of Technology, Brisbane (1999).

\bibitem{rfc6716}
Valin, J.-M., Vos, K., and Terriberry, T.~B., ``{Definition of the Opus Audio
  Codec}.'' RFC 6716 (Proposed Standard) (Sept. 2012).

\bibitem{DaalaWebsite}
``Daala website.'' \url{https://xiph.org/daala/}.

\bibitem{Valin2013}
Valin, J.-M., Maxwell, G., Terriberry, T.~B., and Vos, K., ``High-quality,
  low-delay music coding in the opus codec,'' in [{\em Proc. 135$^{th}$ AES
  Convention}{\nolinebreak\hspace{0.1em}]},  (Oct. 2013).

\bibitem{LG89}
Lookabaugh, T.~D. and Gray, R.~M., ``High-resolution quantization theory and
  the vector quantizer advantage,'' {\em {IEEE} Transactions on Information
  Theory}~{\bf 35},  1020--1033 (Sept. 1989).

\bibitem{Fisher1986}
Fischer, T.~R., ``A pyramid vector quantizer,'' {\em IEEE Trans. on Information
  Theory}~{\bf 32},  568--583 (1986).

\bibitem{Valin2010}
Valin, J.-M., Terriberry, T.~B., Montgomery, C., and Maxwell, G., ``A
  high-quality speech and audio codec with less than 10 ms delay,'' {\em IEEE
  Trans. Audio, Speech and Language Processing}~{\bf 18}(1),  58--67 (2010).

\bibitem{johnston1988transform}
Johnston, J.~D., ``Transform coding of audio signals using perceptual noise
  criteria,'' {\em Selected Areas in Communications, IEEE Journal on}~{\bf
  6}(2),  314--323 (1988).

\bibitem{fletcher1937relation}
Fletcher, H. and Munson, W.~A., ``Relation between loudness and masking,'' {\em
  The Journal of the Acoustical Society of America}~{\bf 9}(1),  78--78 (1937).

\bibitem{Chen2010}
Chen, M.-J. and Bovik, A.~C., ``Fast structural similarity index algorithm,''
  in [{\em Proc. ICASSP}{\nolinebreak\hspace{0.1em}]},   994--997 (march 2010).

\bibitem{DaalaGit}
``Daala git repository.'' \url{https://git.xiph.org/daala.git}.

\end{thebibliography}

\end{document}